\newcommand{\pcmq}{\mbox{cm$^{-2}$}}
\newcommand{\psec}{\mbox{s$^{-1}$}}
\newcommand{\pkev}{\mbox{keV$^{-1}$}}
\newcommand{\funit}{\mbox{ph \pcmq \psec}}
\newcommand{\feunit}{\mbox{ph \pcmq \psec \pkev}}

\def\MeV{\mbox{Me\hspace{-0.1em}V}}
\def\deg{\ensuremath{^\circ}}

\documentclass[a4paper]{spie}
\usepackage{graphicx}

\title{GRI: focusing on the evolving violent Universe}
\author{J\"urgen Kn\"odlseder\supit{a},
        Peter von Ballmoos\supit{a},
	Filippo Frontera\supit{b},
        Angela Bazzano\supit{c},
	Finn Christensen\supit{d},
	Margarida Hernanz\supit{e} and
	Cornelia Wunderer\supit{f}
\skiplinehalf
\supit{a}Centre d'\'Etude Spatiale des Rayonnements, 
         UPS/CNRS, B.P. 44346, 
	 9, avenue du Colonel-Roche, 
	 31028 Toulouse,
	 France;\\
\supit{b}Dipartimento di Fisica, 
         University of Ferrara, 
	 Via Saragat, 1,
	 44100 Ferrara,
	 Italy;\\
\supit{c}INAF-IASF/Rome,
         Via Fosso del cavaliere 100,
	 00133 Roma,
	 Italy;\\
\supit{d}Danish National Space Center,
         Juliane Maries Vej 30, 
	 2100 Copenhagen,
	 Denmark;\\
\supit{e}Institut de Ci\`encies de l'Espai (CSIC-IEEC), 
         Campus Universitat Aut\`onoma de Barcelona, 
         Facultat de Ci\`encies, 
	 Torre C5 - parell - 2a planta, 
	 08193 Bellaterra (Barcelona),
	 Spain;\\
\supit{f}Space Science Laboratory,
         UC Berkeley,
	 7 Gauss Way, Berkeley, CA 94720-7450,
	 USA
}

\authorinfo{Further author information: (Send correspondence to 
J.K.)\\J.K.: E-mail: knodlseder@cesr.fr, Telephone: +33 (0)5 61 55 66 63
\\This paper has been published on behalf of the GRI consortium. 
The GRI consortium is composed of members from the following 
countries (in alphabetical order) 
Belgium,
China,
Denmark, 
France, 
Germany, 
Ireland,
Italy, 
Poland, 
Portugal, 
Russia, 
Spain, 
The Netherlands,
Turkey,
United Kingdom, and
the United States of America.
A complete list of GRI consortium members can be found on
{\tt http://gri.rm.iasf.cnr.it/}}

\begin{document}
\maketitle

\begin{abstract}

The Gamma-Ray Imager (GRI) is a novel mission concept that will provide 
an unprecedented sensitivity leap in the soft gamma-ray domain by using 
for the first time a focusing lens built of Laue diffracting crystals. 
The lens will cover an energy band from 200 - 1300 keV with an effective 
area reaching 600 cm$^2$. 
It will be complemented by a single reflection multilayer coated 
mirror, extending the GRI energy band into the hard X-ray regime, down to 
$\sim$10 keV.
The concentrated photons will be collected by a position sensitive 
pixelised CZT stack detector.
We estimate continuum sensitivities of better than $10^{-7}$ \feunit\
for a 100 ks exposure; the narrow line sensitivity will be better than 
$3 \times 10^{-6}$ \funit\ for the same integration time. 
As focusing instrument, GRI will have an angular resolution of better than 
30 arcsec within a field of view of roughly 5 arcmin - an unprecedented 
achievement in the gamma-ray domain.
Owing to the large focal length of 100 m of the lens and the mirror, 
the optics and detector will be placed on two separate spacecrafts flying 
in formation in a high elliptical orbit. 
R\&D work to enable the lens focusing technology and to develop the required 
focal plane detector is currently underway, financed by ASI, CNES, ESA, 
and the Spanish Ministery of Education and Science. 
The GRI mission is proposed as class M mission for ESAÕs Cosmic Vision 
2015-2025 program. 
GRI will allow studies of particle acceleration processes and explosion 
physics in unprecedented detail, providing essential clues on the innermost 
nature of the most violent and most energetic processes in the Universe.
\end{abstract}
    

\section{INTRODUCTION}

Following 4 years of successful operations, INTEGRAL has 
significantly changed our vision of the gamma-ray sky 
\cite{winkler06}.
The telescopes aboard the satellite have revealed hundreds of sources 
of different types, new classes of objects, extraordinary and 
puzzling views of antimatter annihilation in our Galaxy, and 
fingerprints of recent nucleosynthesis processes.
With the wide fields of view of the IBIS and SPI telescopes, INTEGRAL is 
an exploratory-type mission \cite{winkler03}
that allows extensive surveys of the hard 
X-ray and soft gamma-ray sky, providing a census of the source 
populations and first-ever all-sky maps in this interesting energy range.
The good health of the instruments allows continuing the exploration during 
the upcoming years, enabling INTEGRAL to provide the most complete and 
detailed survey ever, which will be a landmark for the discipline throughout 
the next decades.

Based on the INTEGRAL discoveries and achievements, there is now a growing 
need to perform more focused studies of the observed phenomena.
High-sensitivity investigations of point sources, such as compact 
objects, pulsars, and active galactic nuclei, should help to uncover 
their yet poorly understood emission mechanisms.
A deep survey of the galactic bulge region with sufficiently 
high-angular resolution should shed light on the still mysterious 
source of positrons.
And a sensitivity leap in the domain of gamma-ray lines should allow 
the detection of nucleosynthesis products in individual supernova 
events, providing direct insights into the physics of the exploding 
stars.

\begin{figure}[!t]
\begin{center}
\includegraphics*[width=17cm]{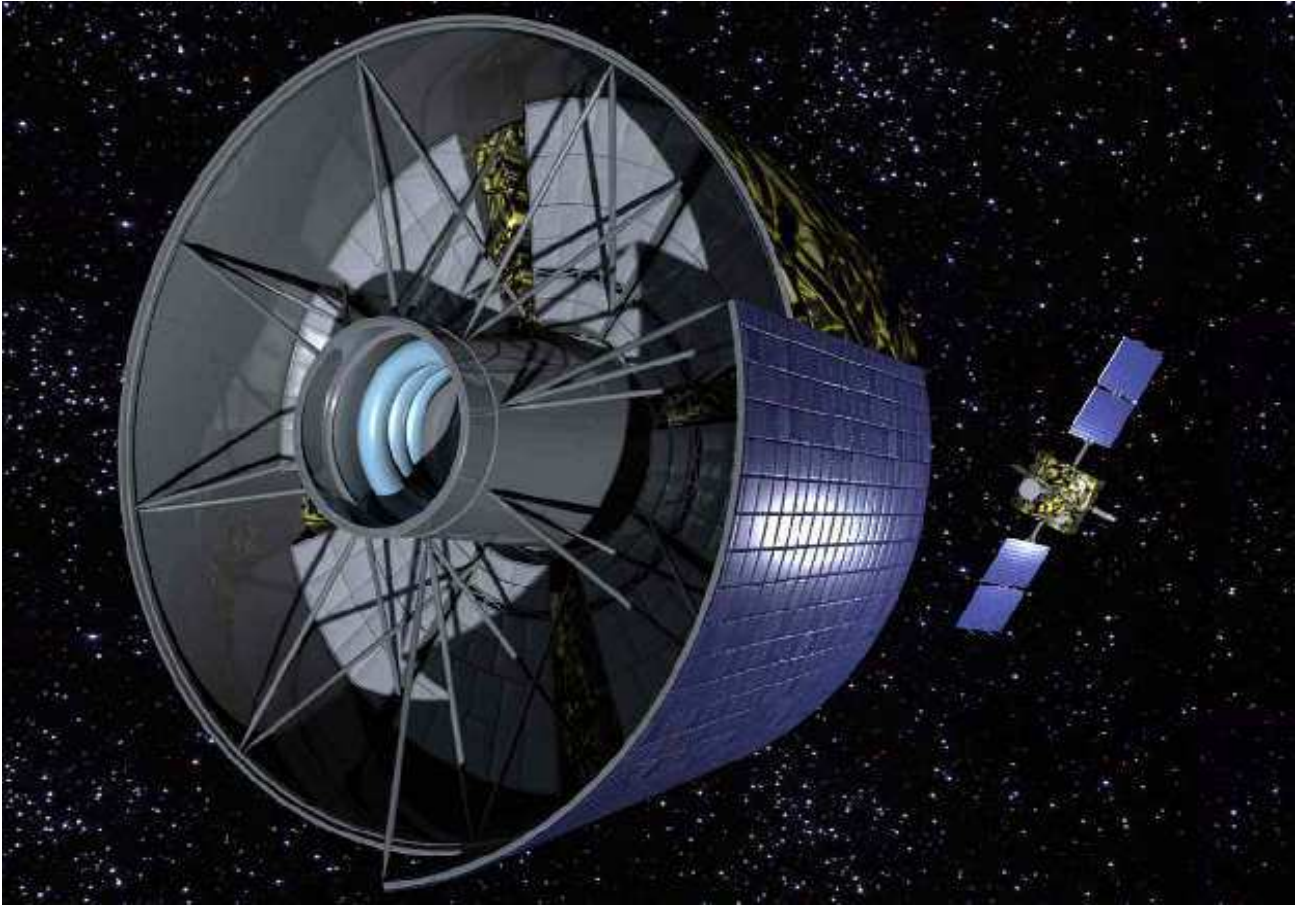}
\end{center}
\caption{\label{figure1}
The GRI mission deployed in space.
GRI is composed of two spacecrafts flying in formation at a distance 
of 100~m.
The Optics Spacecraft (front) carries the Laue crystal lens and 
the multilayer coated single reflection mirror payloads that 
concentrate incoming hard X-rays and soft gamma-rays onto a focal 
spot situated 100~m behind the spacecraft.
The Detector Spacecraft (back), carrying the detector payload, is placed 
at the location of the focal spot to collected the concentrated photons.
}
\end{figure}

Technological advances in the past years in the domain of gamma-ray 
focusing using Laue diffraction have paved the way towards a new 
gamma-ray mission that can fulfil these requirements.
Laboratory work and balloon campaigns have provided the 
proof-of-principle for using Laue lenses as focusing devices in  
gamma-ray telescopes \cite{ballmoos04, halloin04,barriere06,frontera06}, 
and concept studies by CNES and ESA have demonstrated that such an 
instrument is technically feasible and affordable 
\cite{duchon05, brown05}.
Complemented by a hard X-ray telescope based on a single-reflection 
multilayer coated concentrator, a broad-band energy coverage can be achieved 
that allows detailed studies of astrophysical sources at 
unprecedented sensitivity and angular resolution, from $\sim10$
keV up to at least 1 \MeV.

Bringing our scientific requirements into the context of these 
technological achievements, we started a common effort to define the
scenario for a future gamma-ray mission that we baptised the 
{\em Gamma-Ray Imager} (GRI).
The GRI mission fits well into the framework of ESA's Cosmic Vision 
2015-2025 planning, and it will provide a perfect successor for 
INTEGRAL in that it will considerably deepen the study of the 
phenomena unveiled by the observatory.
In response to the first Announcement of Opportunity of the Cosmic 
Vision program, a class M mission proposal has been submitted recently 
to ESA.
Figure \ref{figure1} shows an artists view of the GRI mission 
deployed in space.
The image is based on the GRI mission concept study conducted by Thales 
Alenia Space in the context of the GRI Cosmic Vision proposal.

\section{GRI SCIENCE}
\label{sec:science}

Many fundamental questions of modern astrophysics are related to the 
extremes of our Universe: extreme energies that drive powerful stellar 
explosions and accelerate particles to macroscopic energies, extreme 
densities that modify the laws of physics around the most compact objects 
known, and extreme fields that influence the matter in a way that is 
unexplorable on Earth.
The Gamma-Ray Imager (GRI) will explore these extremes via focused 
observations in the hard X-ray and soft gamma-ray bands. 
For the first time ever in this domain, focusing technologies will be 
employed to concentrate high-energy photons onto a small focal spot. 
Focusing will bring the long awaited sensitivity leap, lifting the veil 
of the extreme Universe.
The main science questions that GRI will address and that drive the 
mission design are:
\begin{itemize}
\item {\bf How do supernovae explode?}
      GRI will measure gamma-ray line and continuum light curves and 
      line shapes of radioactive decay products in thermonuclear 
      explosions.
      This will unveil the Type Ia supernova explosion mechanisms and 
      provides the key to understand the variety of the SN~Ia phenomena.
      Furthermore, the GRI observations will provide a calibration of 
      the SN~Ia luminosities as required for accurate measurements of
      the cosmic acceleration.
\item {\bf What is the origin of the soft gamma-ray background radiation?}
      GRI will determine the spectral properties of accretion and jet 
      dominated AGN (continuum shapes, cut-off energies).
      The characterisation of AGN high-energy spectra will constrain 
      their contributions to the cosmic background radiation, and will 
      probe their radiation physics.
\item {\bf What links jet ejection to accretion in black hole and neutron 
       star systems?}
      GRI will measure spectral properties during state transitions, 
      determine the polarization of the emission, and search for 
      annihilation signatures.
      These measurements will reveal how the state transitions are 
      triggered.
      Furthermore they will constrain the system geometries, and 
      determine the nature of the emitting hot plasma.
\item {\bf How are particles accelerated to extreme energies in the 
       strongest magnetic fields?}
      GRI will perform phase resolved spectroscopy of pulsars and
      determine the polarization of the high-energy emission, 
      constraining thus the geometry of pulsar magnetosphere.
      Moreover, GRI will measure the spectral properties of magnetars, 
      determining their acceleration physics and probing QED effects 
      in the strongest known magnetic fields.
\end{itemize}

In addition to these main objectives, GRI will also detect radioactive 
decay products in nearby novae or young supernova remnants, constraining 
their nucleosynthesis and stellar physics. 
Positron-electron annihilation signatures constitute an important 
diagnostics tool of particle populations that GRI will search for in a 
variety of objects such as galactic compact objects, pulsars, supernovae 
and supernova remnants. 
Particle acceleration phenomena can further be studied in supernova remnants, 
galaxy clusters, and gamma-ray bursts. 
And GRI also has the potential to constrain or even detect the products 
of dark matter annihilation and/or decay.

\begin{figure}[!t]
\begin{center}
\includegraphics*[height=4.35cm]{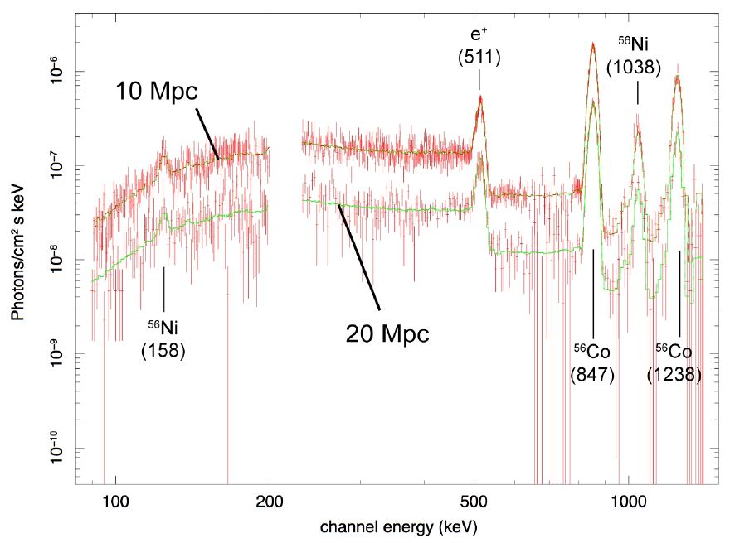}
\hfill
\includegraphics*[height=4.35cm]{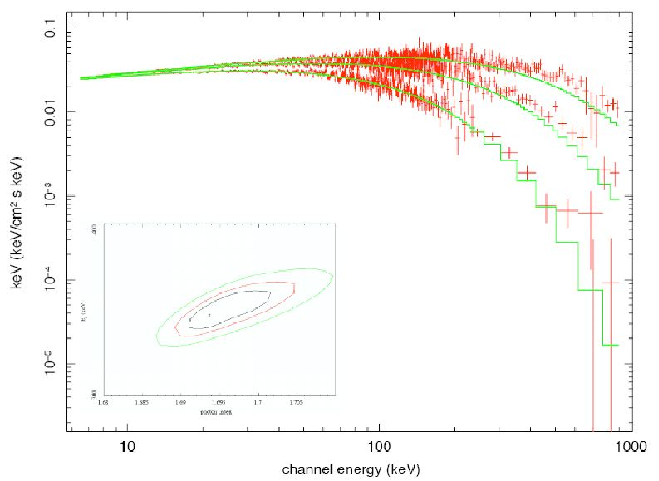}
\hfill
\includegraphics*[height=4.35cm]{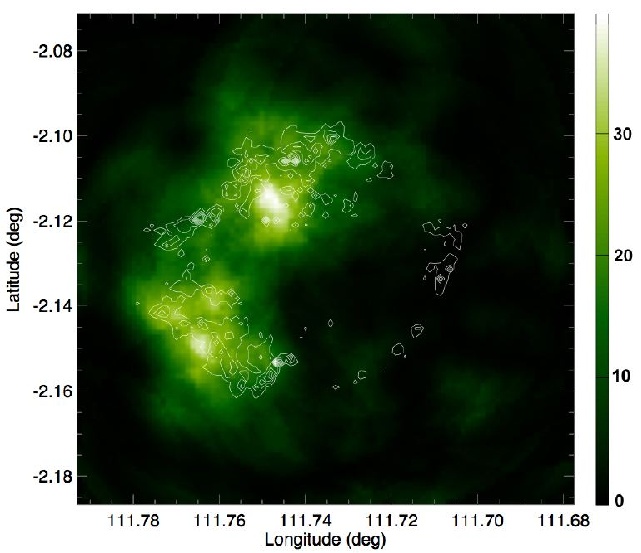}
\end{center}
\caption{\label{fig:science}
GRI science simulations:
{\it Left panel:} Type~Ia supernova spectra obtained for events at 
10 Mpc and at 20 Mpc, respectively.
{\it Mid panel:} AGN spectra for different cut-off energies.
{\it Right panel:} Image of the (assumed) spatial distribution of 
$^{44}$Ti in the Cas A supernova remnant.
}
\end{figure}

Figure \ref{fig:science} shows three science simulations for GRI.
The left panel shows the observed spectrum for thermonuclear 
supernovae (SN~Ia) at distances of 10 Mpc and 20 Mpc for an observing 
time of 1 Ms.
A delayed detonation model has been assumed and the spectra have been 
simulated for 90 days after the explosion where the gamma-ray line intensities 
are at maximum. 
The most important lines arise from the radioactive decay of $^{56}$Ni and 
$^{56}$Co, and from electron-positron annihilation, and all of them 
can be clearly distinguished by GRI.
GRI will measure such spectra for about 5 events per year.

The mid panel shows Seyfert spectra with different cut-off energies for a 
flux of 3 mCrab in the 2-10 keV energy range. 
The simulation clearly demonstrates the GRI capability to measure the high 
energy cut-off up to highest energies with unprecedented accuracy in a 100 ks 
observation, crucial to determine the contribution of these objects 
to the extragalactic diffuse background.

The right panel shows the image that GRI will obtain for a 1 Ms observation 
of the Cas A supernova remnant in the 68 keV decay line of $^{44}$Ti. 
The observation has been simulated by assuming that the $^{44}$Ti is 
distributed similar to Si, as mapped by Chandra observations (white contours). 
Clearly, the GRI observation will allow to localise the radioactive material 
in the supernova remnant, providing important insights into the dynamics 
and symmetry of the explosion that took place 320 years ago!

\section{GRI MISSION}

\subsection{Mission profile}

\begin{figure}[!h]
\begin{center}
\includegraphics*[height=6cm]{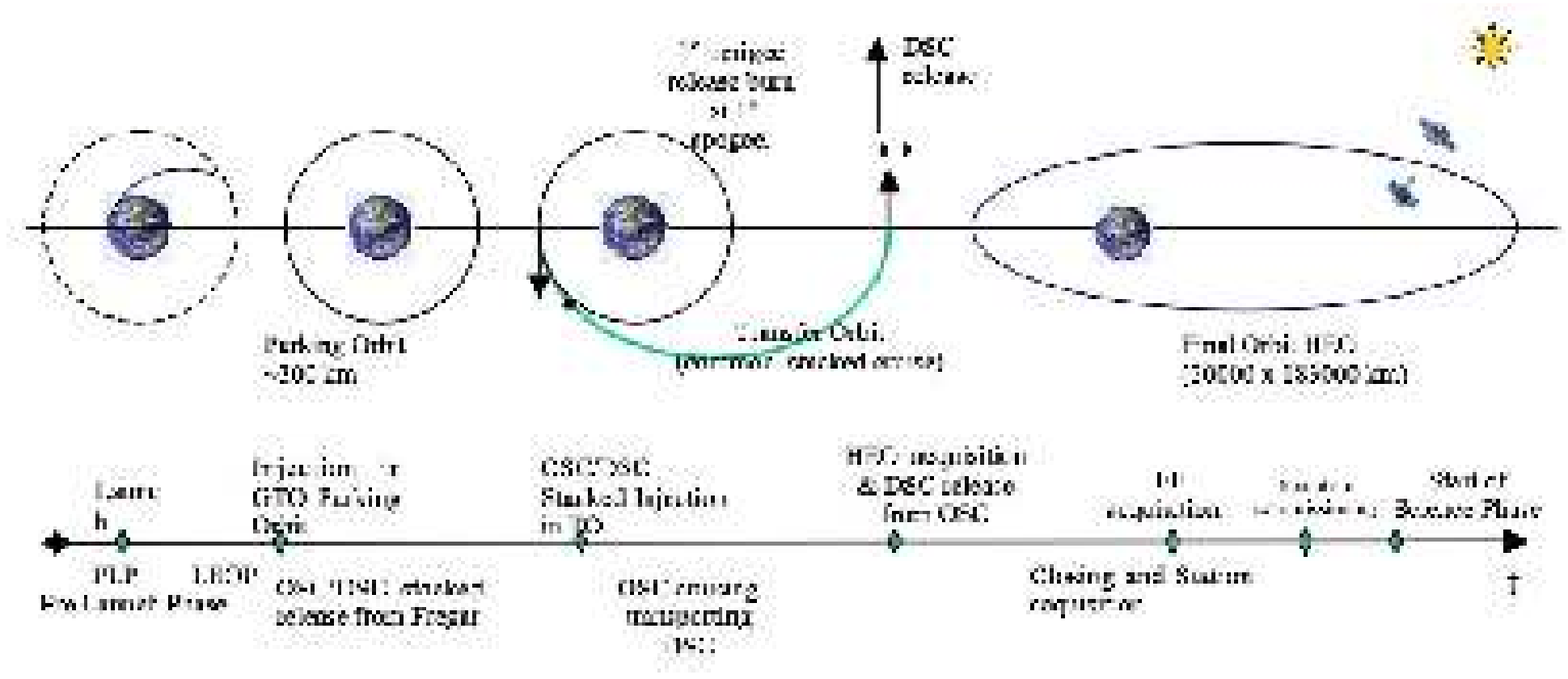}
\includegraphics*[height=6cm]{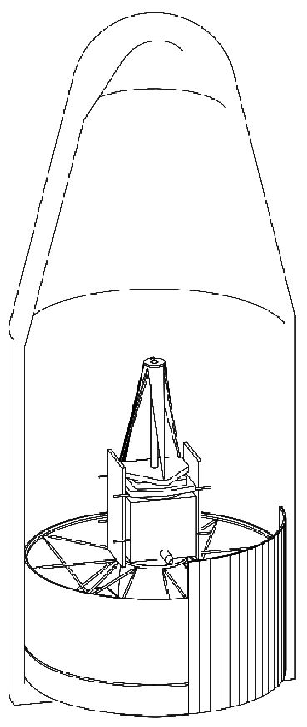}
\end{center}
\caption{\label{fig:profile}
GRI mission profile ({\it left}) and launch configuration ({\it right}).
}
\end{figure}

The GRI mission is composed of two spacecrafts that will be launched by 
a single Soyuz Fregate-2B launch from Kourou. 
The optics spacecraft (OSC) carries a multilayer coated single-reflection 
mirror and a Laue crystal lens that focus incoming gamma-rays onto a focal 
spot situated at a distance of 100~m. 
The detector spacecraft (DSC) carries a position sensitive detector, and 
will be placed in the focal spot to collect the concentrated gamma-rays. 
Figure \ref{fig:profile} illustrates the GRI mission profile. 
Both satellites will be launched stacked in a circular 300~km parking orbit 
(a similar to the configuration that has been proposed and validated for the 
Simbol-X and XEUS missions). 
The heavier OSC will be placed at the bottom, and will be supported by the 
Soyuz-Fregate standard adapter (1194-SF). 
The lighter DSC will be placed on top of the OSC. 
For this purpose the OSC is equipped with a support cone with a DSC interface 
diameter of 937 mm. 
With an overall mass budget of 2340 kg, and including the necessary S/C 
adapter, the GRI total launch mass (including margins) amounts to 2327 kg, 
leaving a performance margin of 13 kg with respect to the launcher performances 

Using the OSC propulsion system, the satellites are then injected stacked 
into a transfer orbit to reach a final highly eccentric orbit (HEO).
A HEO operational orbit with perigee between 15000 - 20000 km provides 
the best conditions for mission maintenance, formation flying control, 
ground station link as well as scientific efficiency. 
The apogee of the orbit will be at 183000 km, leading to a revolution 
period of 4 days 
(the orbit is identical to that proposed and validated for the Simbol-X 
mission).
Once in final orbit, both satellites will be separated, acquire the Sun, 
and put into formation. 
After a scientific commissioning phase, a science phase of 3 + 1 years will 
follow.

\subsection{Optics Spacecraft}

\begin{figure}[!t]
\begin{center}
\includegraphics*[width=5cm]{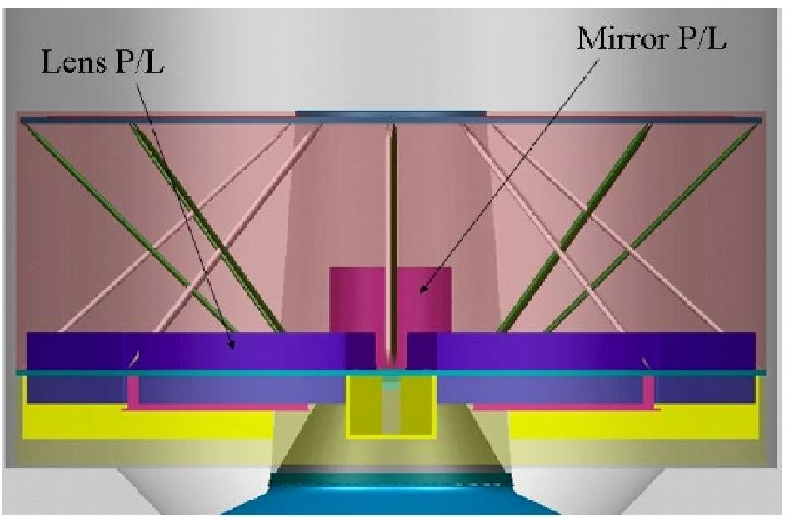}
\includegraphics*[width=5cm]{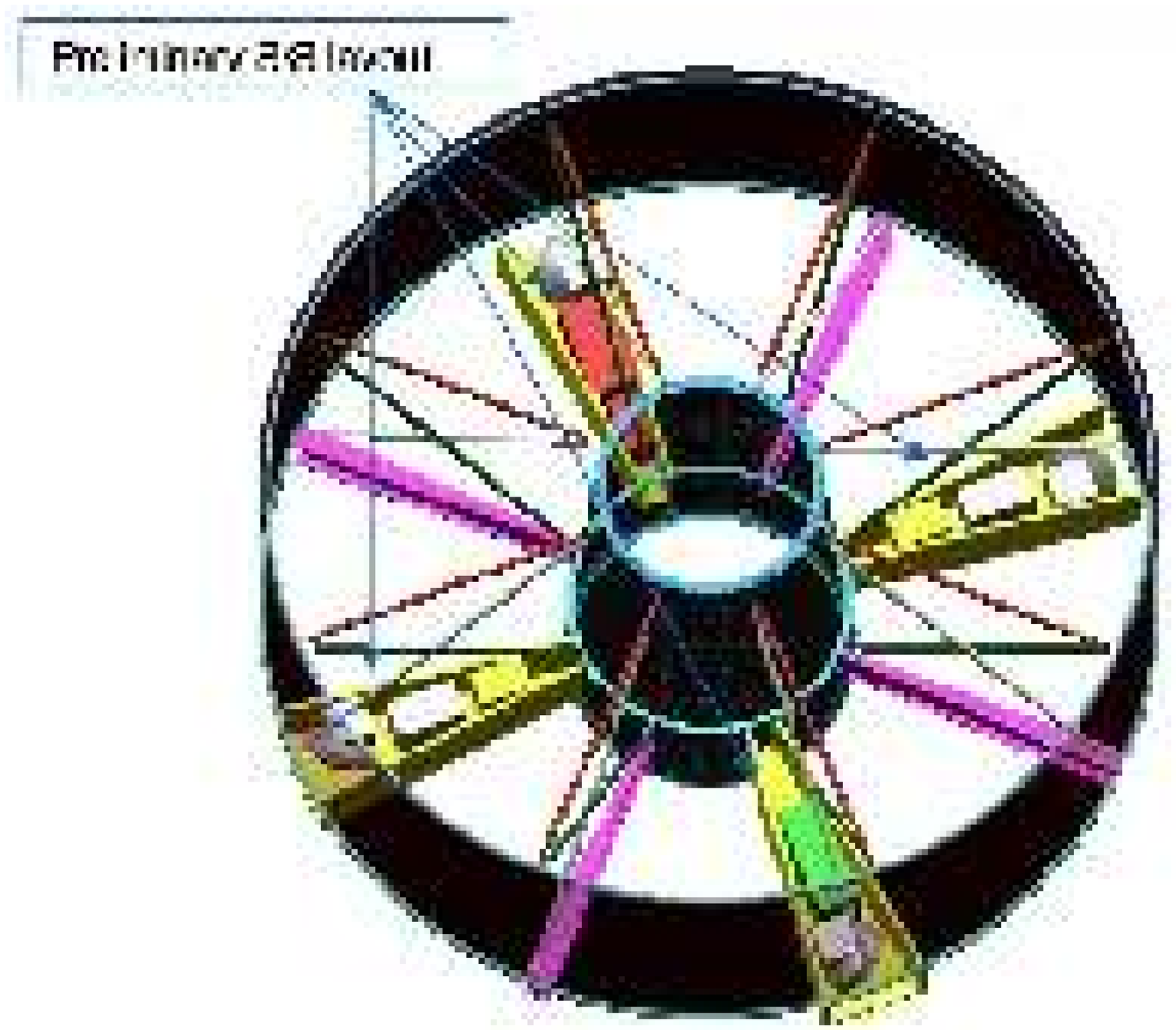}
\includegraphics*[width=6cm]{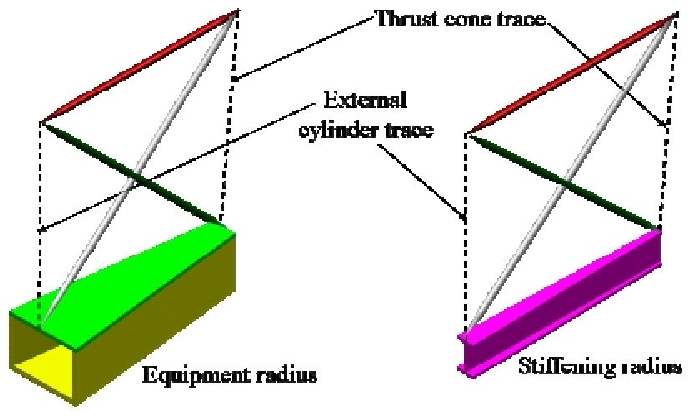}
\end{center}
\caption{\label{fig:osc}
Overview over the OSC design.
From left to right: OSC configuration, preliminary sub-systems layout, and 
radius.
}
\end{figure}

For the optics spacecraft (OSC) we propose a specific satellite development 
that aims in the maximization of the surface area usable for the collection 
of gamma-rays, while satisfying the launch and operational constraints of 
the GRI mission.
A specific OSC design study in the context of the Cosmic Vision 
proposal has been performed by TAS-I Turin.
While the OSC requires a dedicated solution, the S/C can be equipped with 
standard equipment at sub-system level, with advantages in terms of cost 
and reliability. 
The OSC preliminary design concept is based on XEUS heritage for the wheel 
geometry and truss support structure. 
It consists of a thrust cone, four equipment radius, four stiffening radius, 
eight sets of reinforcement rods and an external cylindrically drum to 
close the load flux (cf. Fig.~\ref{fig:osc}). 
The thrust cone reaches from the launcher interface ring through the OSC 
to the DSC interface ring, allowing for the launch of both S/C in a stacked 
configuration. 
The four equipment radius host the main S/C sub-systems. 

In order to have the electrical power generation compatible with the 
constraints of the OSC geometry, a system based on solar photovoltaic 
cells mounted on the OSC drum has been considered. 
The cells have been sized for a Sun aspect angles of $\pm20\deg$. 
To allow also larger solar aspect angles, required for the survey of 
supernovae, a specific mode has been considered where all systems are 
powered from battery with autonomy of 2-3 hours. 
In this case, the solar aspect angle can be extended to $\pm80\deg$.

\subsection{Detector Spacecraft}

For the detector spacecraft (DSC) we propose the utilization of the 
TAS-F PROTEUS satellite, a multi-purpose platform developed for the so-called 
mini satellites family that is flight proven with the successful launch 
of the JASON-1 and COROT satellites. 
The use of the qualified PROTEUS platform for GRI leads to reduced risks, 
cost minimization, and simplifies the qualification and AIT sequence. 
The robust and reliable PROTEUS structure is composed of an aluminium rod 
assembly forming a cube (1 m long) stiffened by sandwich panels. 
It acts as primary structure and guarantees the transfer of main loads from 
payload to launcher through a fully machined aluminium frame. 
The PROTEUS structure was qualified on former programs for payloads up to 
300 kg and is thus fully compliant with the payload mass of 120 kg. 
The payload telemetry of 120 kbps is compatible with the PROTEUS standard 
capability.

\begin{figure}[!t]
\begin{center}
\includegraphics*[width=12cm]{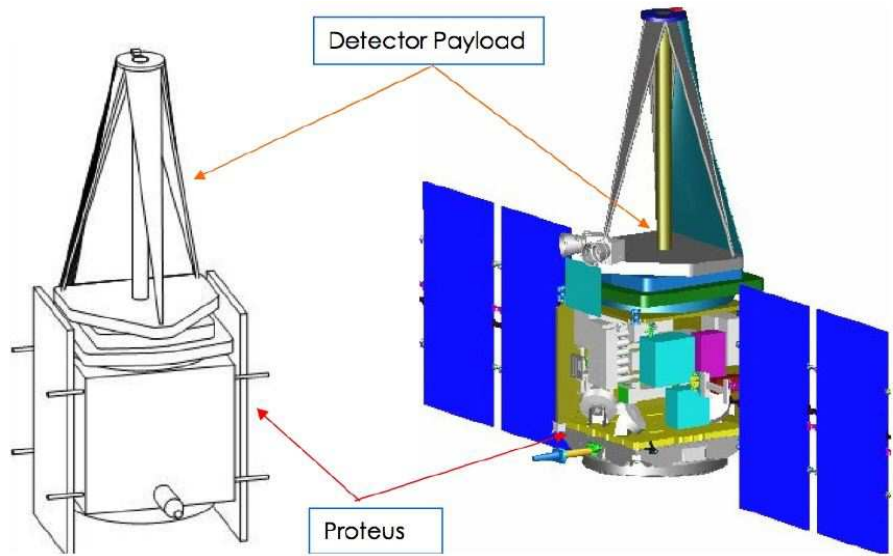}
\end{center}
\caption{\label{fig:dsc}
DSC with integrated payload.
}
\end{figure}

\subsection{Formation flying metrology}

The GRI telescope is split into two satellites: the OSC carrying the lens 
and the mirror and the DSC maintains actively an accurate formation flight 
(position and orientation) with respect to the OSC at the focal length of 
100 m. 
For the observation, the formation flight maintains an inertial pointing 
towards the source, requiring an absolute accuracy of 10 arcsec. 
The OSC is in charge of controlling its attitude using Reaction Wheels, 
which are unloaded by the chemical propulsion system. 
The precision of the star trackers will allow to determine the effective 
OSC pointing to an accuracy of 3 arcsec.

The GRI formation flying control is based on the relative position 
measurement performed by the metrology devices. 
On the basis of the measured relative OSC/DSC position, the DSC onboard 
computer elaborates thanks to dedicated software the alignment and attitude 
errors to be compensated by the means of cold gas thrusters.

Reorientation to a new science target is performed by slewing the OSC and 
translating and slewing the DSC in order to maintain its focal position. 
Typically, $20\deg$ rotations are performed in roughly 2 hours, with a cold 
gas reserve for 800 target changes.

\begin{figure}[!t]
\begin{center}
\includegraphics*[height=3.4cm]{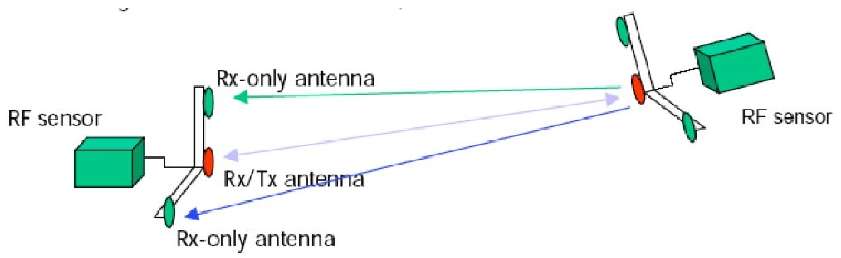}
\includegraphics*[height=3.4cm]{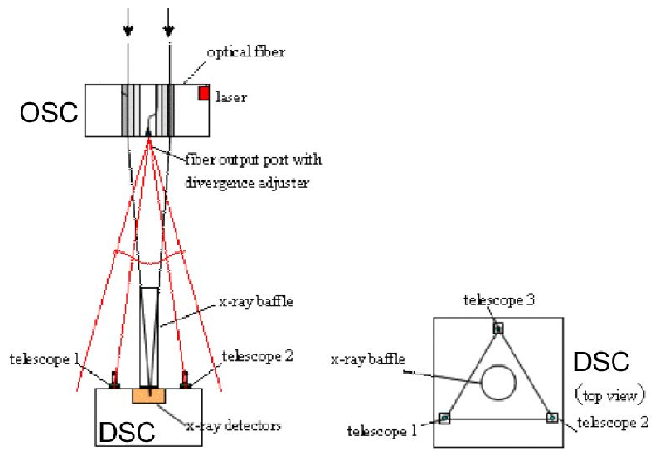}
\end{center}
\caption{\label{fig:formation}
Formation flying metrology of GRI.
{\it Left panel:} RF metrology and mounting scheme.
{\it Right panel:} HDOM metrology and mounting scheme.
}
\end{figure}

Two types of metrology systems are foreseen: RF metrology and optical 
metrology. 
The proposed RF sensor has been designed by TAS-F and belongs to the coarse 
metrology sensors family, providing longitudinal ranging measurements and 
bearing angles for azimuth and elevation determination. 
The RF metrology offers also an inter-satellite data-link (ISL), which will 
be used for HK data transfer from OSC to DSC, OSC commanding and time 
coordination (bi-directional throughput of 10 kbps). 
The following equipment is foreseen: On the OSC, a total of 6 antennae 
are installed, where 3 are facing the DSC in formation flying condition, 
and 3 are facing opposite to provide $4\pi$ coverage for safety reasons. 
Symmetrically, 6 antennae are installed on the DSC, 3 facing the OSC and 
3 on the opposite side. 
Each set of 3 antennae are placed in a L shape configuration 
(Fig.~\ref{fig:formation}).

A Hexa-Dimensional Optical Metrology (HDOM) system provides the 3D position 
and 3D orientation of the DSC relative to the reference frame of the OSC. 
The HDOM, developed by TAS-I, has been considered for GRI because of the 
robustness in luminous environments, allowing for very wide solar 
aspect angles of $\pm80\deg$ as required for the supernova survey. 
The HDOM system elements are split between the two satellites
(Fig.~\ref{fig:formation}). 
The relatively weak requirements of the GRI mission leave generous 
performance margins.

\section{GRI PAYLOADS}

The GRI payload is composed of a Laue crystal lens, a multilayer-coated 
single-reflection mirror and a focal plane detector. 
The lens and the mirror are placed on the OSC while the focal plane detector 
is placed on the detector spacecraft DSC.

\subsection{Laue lens payload}

The Laue lens payload consists of a broad-band gamma-ray lens based on the 
principle of Laue diffraction of photons in mosaic crystals. 
Each crystal can be considered as a little mirror which deviates gamma-rays 
through Bragg reflection from the incident beam onto a focal spot. 
Although the Bragg relation $2d \sin \theta = nhc / E$ implies that only a 
single energy E (and its multiples) can be diffracted by a given crystal, 
the mosaic spread $\Delta\theta$ that occurs in each crystal leads to an 
energy spread $\Delta E \propto (\Delta\theta)E^2$ ($d$ is the crystal lattice 
spacing, $\theta$ the Bragg angle, $n$ the diffraction order, $h$ the Planck 
constant, $c$ the speed of light and $E$ the energy of the incident photon). 

Placing the crystals on concentric rings around an optical axis, and the 
careful selection of the inclination angle on each of the rings, allows then 
to build a broad-band gamma-ray lens that has continuous energy coverage over 
a specified band.
Since large energies $E_1$ imply smaller diffraction angles $\theta$, 
crystals diffracting large energies are located on the inner rings of the 
lens. 
Conversely, smaller energies $E_2$ imply larger diffraction angles and 
consequently the corresponding crystals are located on the outer rings.

\begin{figure}[!ht]
\begin{center}
\includegraphics*[height=5cm]{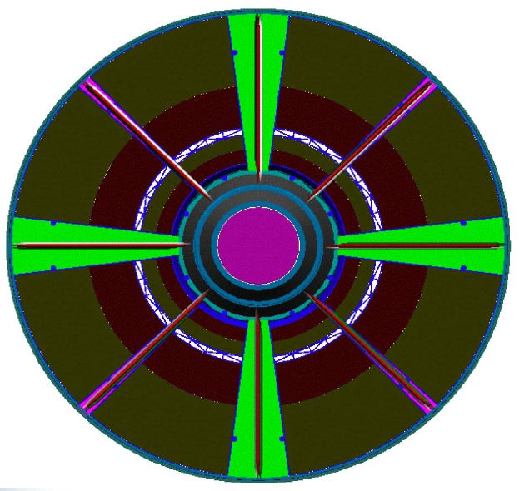}
\includegraphics*[height=5cm]{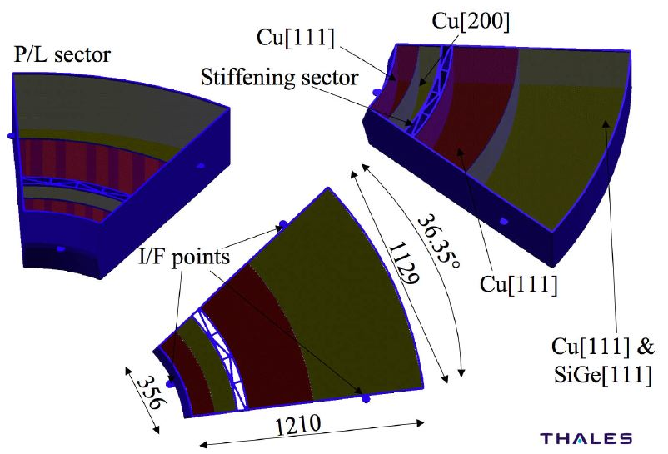}
\end{center}
\caption{\label{fig:lens}
Overview over the lens payload.
}
\end{figure}

The lens payload consists of a set of 8 petals, each of which is interfacing 
at 3 points with the OSC (one point on the thrust cone, one on an equipment 
radius and one on a stiffening radius). 
This concept (illustrated in Fig.~\ref{fig:lens}) is inspired from the XEUS 
MSC design, and allows a good lens payload modularity, with advantages in 
terms of feasibility, testability and production. 
Each petal has an opening angle of $36.4\deg$ and a length of 121 cm, and 
is composed of a SiC frame and support structure.
Each petal will be divided into 22 sub-sectors that hold the crystal modules. 
Two options for the assembly of crystals on these modules are currently 
under study in dedicated industrial studies (CNES funded: contractor TAS-F 
Cannes; ESA funded: contractor TAS-I Milano).

\subsection{Multilayer coated mirror}

The payload consists of a multilayer coated single reflection mirror that 
acts as a concentrator for hard X-rays comprised in the energy band from 
10 to 300 keV. 
We propose to use the high performance Si-pore optics developed by ESA 
within the XEUS mission concept to realize a light-weight mirror substrate
\cite{beijersbergen04}. 
Onto this substrate, depth graded multilayer coatings will be applied which 
act as broad band Bragg reflectors
\cite{christensen06}. 
The basic principle of the mirror is very analogous to the principle used 
for the Laue lens payload, making use of the Bragg relation. 
Instead of using crystalline planes, however, the mirror employs depth 
graded multilayer coatings.

\begin{figure}[!h]
\begin{center}
\includegraphics*[height=5cm]{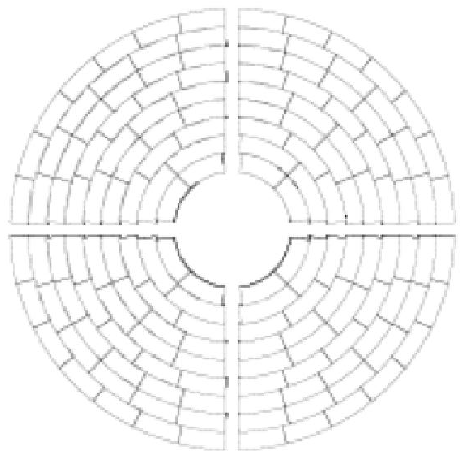}
\includegraphics*[height=5cm]{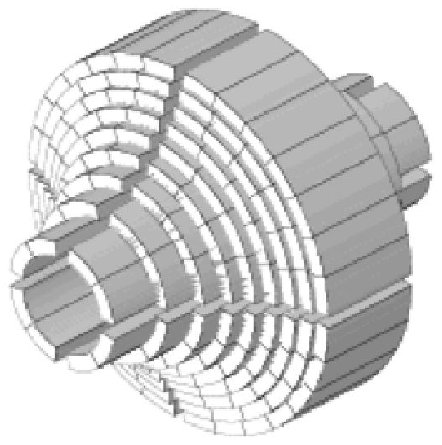}
\end{center}
\caption{\label{fig:mirror}
Schematic representation of the mirror payload.
}
\end{figure}

The mirror design is based on nested pore optics similar to that proposed 
for XEUS. 
With respect to the XEUS mission, however, the GRI mirror is significantly 
simplified. 
Firstly, instead of an approximated Wolter geometry only a single reflection 
is used, leading to a simple concentrator with relaxed alignment constraints. 
Secondly, the mirror dimensions are much smaller than those foreseen for 
XEUS, leading to a simplified petal design providing less obscuration and 
simplified mounting. 
Thirdly, the imaging capabilities are relaxed to 10 arcsec instead of the 
XEUS goal of 2 arcsec, relaxing further the precision requirements for 
the optics.

The mirror module consists of 86 petal units. 
Each of these X-ray Optical Units (XOUÕs) needs to be aligned independently 
and is mounted together by special CeSiC brackets allowing for this alignment. 
Due to the fixed spacing of the Si wafers the inner shells need to be 
longer to make an effective use of the aperture. 
It is thus required to align several XOU's to the required precision along 
the optical axis. 
This petal structure is schematically shown in Fig.~\ref{fig:mirror}.

\subsection{Detector}

As baseline for the GRI focal plane detector, we propose a pixel CZT 
detector operating at room temperature that combines high detection 
efficiency, good spatial and good spectral resolution
\cite{caroli06}. 
Instead of CZT crystals the usage of CdTe crystals could be envisioned 
for more effective signal compensation or improved spectroscopic 
performance, e.g. by the use of p-i-n contacts. 
Alternatively, a cooled high-purity Ge strip detector could provide an 
improved spectral resolution and better background rejection capabilities 
that potentially increase the scientific performances of GRI
\cite{wunderer06}. 
However, the Ge detector option implies active cooling and eventually 
detector annealing cycles that increase the complexity of the payload. 

\begin{figure}[!t]
\begin{center}
\includegraphics*[height=9cm]{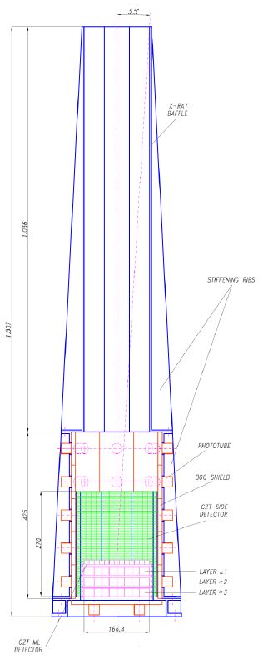}
\includegraphics*[height=9cm]{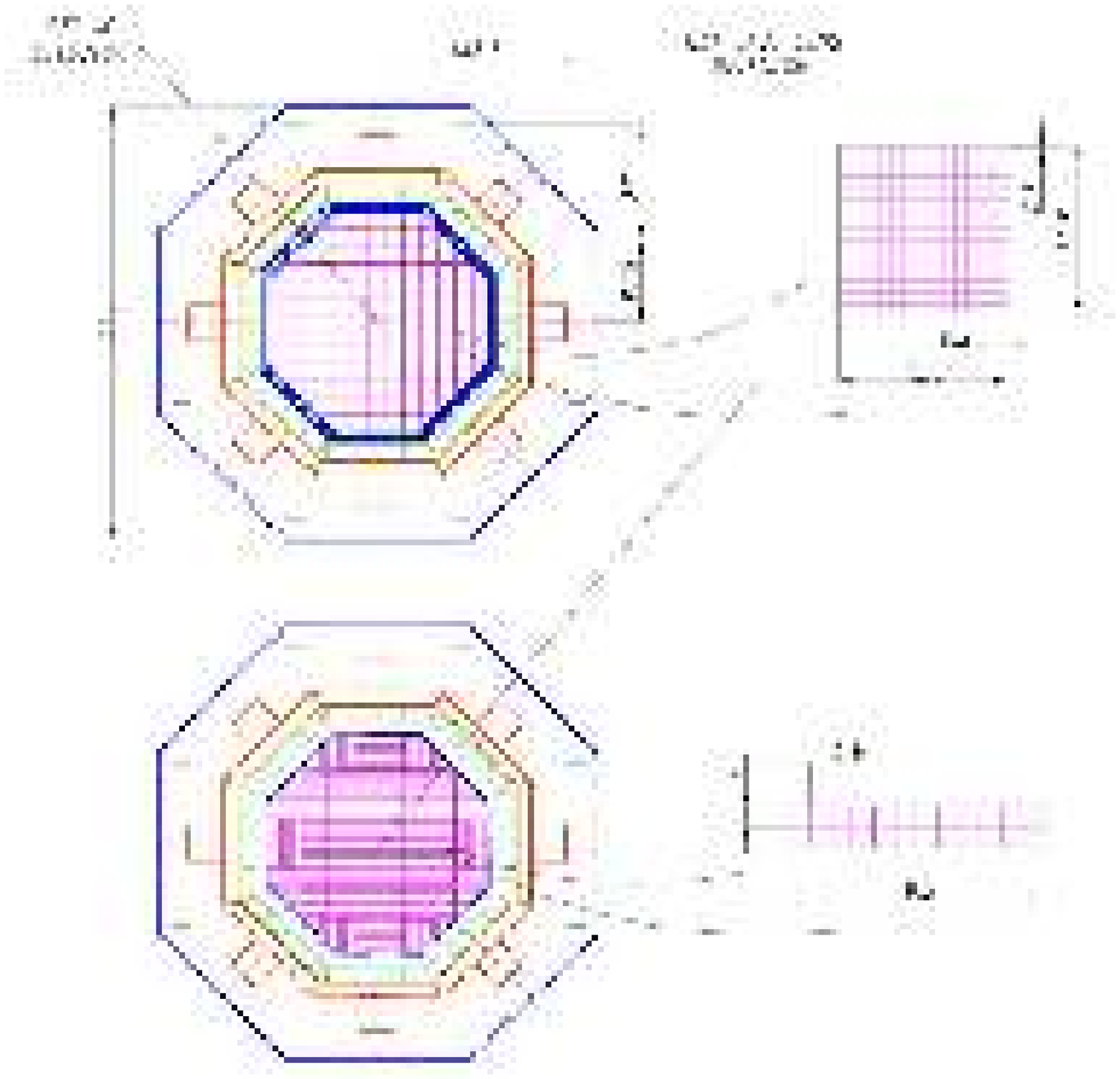}
\end{center}
\caption{\label{fig:detector}
Schematic representation of the detector payload.
}
\end{figure}

The GRI focal plane baseline detector is based on a position sensitive 
spectrometer made of 4 stacked CZT layers, surrounded by CZT side walls. 
The stack is composed of a 5 mm thick top layer, optimized for 
photoelectric absorption in the 10-250 keV band, and three bottom layers, 
each one 20 mm thick, to grant a total detection efficiency of better 
than 75\% for photons below 1 MeV. 
In our preliminary design, the top layer is composed of 129 CZT crystals 
divided into 33024 pixels of $0.8 \times 0.8$ mm$^2$ in size. 
Each of the (identical) bottom layers is composed of 688 CZT crystals 
divided into 8384 pixels of $1.6 \times 1.6$ mm$^2$ in size. 
The crystals of the bottom layers are placed vertically, so that the 
interaction path for the photons amounts to 20 mm within a single crystal. 
In contrast to the top layer, that is biased in planar field configuration, 
the crystals of the bottom layers are biased in planar transverse field 
(PTF) configuration, resulting in small charge collection lengths of 1.6 mm, 
granting a good spectral resolution. 
The detector stack is directly exposed to the source photons focused by 
the lens and mirror optics, while the side wall detectors are used to 
collect scattered photons from the primary beam in order to maximize the 
full energy absorption efficiency. 
In our preliminary design, the side walls are made of 1344 CZT crystals 
divided into 4032 pixels of $6.4 \times 6.4$ mm$^2$ in size. 
The side wall has a thickness of 10 mm, and the crystals are again biased 
in PTF configuration. 
The entire detector is surrounded by a segmented veto shield made of 44 
BGO modules (thickness 10 mm), which are read out through optically 
coupled new generation photomultiplier tubes (PMTs). 
On top of the detector, a baffle made of a sandwich of materials with 
decreasing Z (e.g. W, Sn, Cu) will allow an efficient reduction of the 
diffuse cosmic background for energies up to 100 keV. 

Figure \ref{fig:detector} shows the preliminary design of the GRI detector 
payload. 
The high detector segmentation will allow on the one hand for precise 
localisation of the photon interaction locations and for Compton polarimetry, 
and on the other hand for effective background rejection exploiting 
Compton kinematics event reconstruction techniques. 
Our preliminary detector design consists of 62208 individual CZT pixels 
and 44 PMTs. 

Each CZT pixel, grouped in homogeneous families according to position, will 
be read out individually using a low-power charge pre-amplifier coupled 
through a multiplexing unit to a flash ADC. 
This functionality should be integrated into dedicated multichannel ASICs. 
The ASIC outputs will be collected in the Instrument Control Unit (ICU) 
where a FPGA based digital processing unit provides the event 
characterisation (total energy deposit, interaction position, time tagging 
and coincidence analysis for Compton kinematics reconstructions and 
polarimetry analysis). 
The onboard event processing will lead to a significant data reduction that 
allows to minimize the telemetry downlink rate.

\subsection{Performance assessment}

\begin{figure}[!t]
\begin{center}
\includegraphics*[width=8cm]{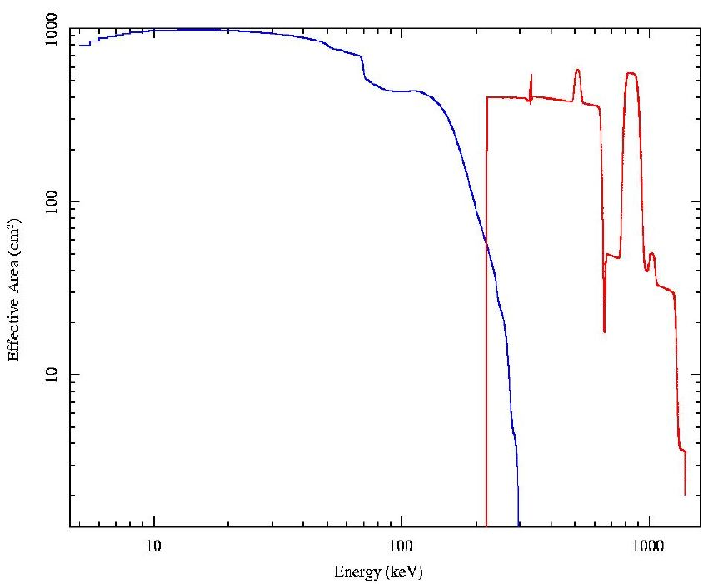}
\includegraphics*[width=8cm]{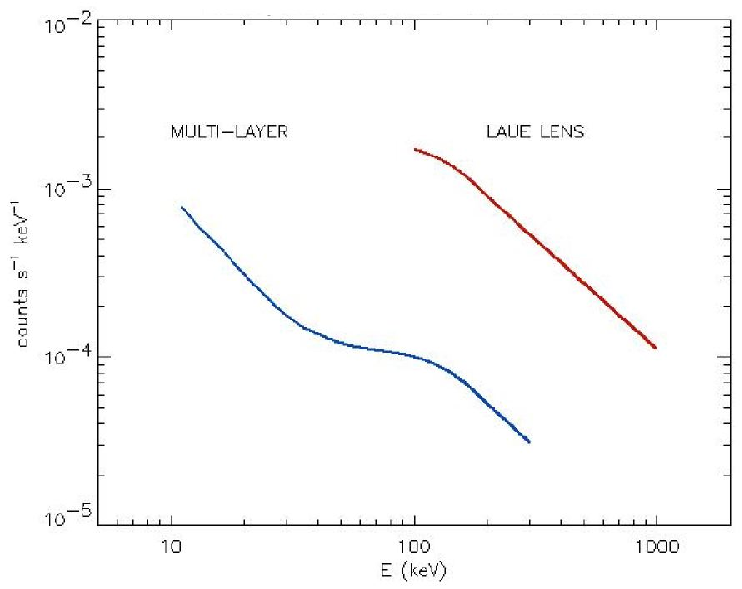}
\includegraphics*[width=8cm]{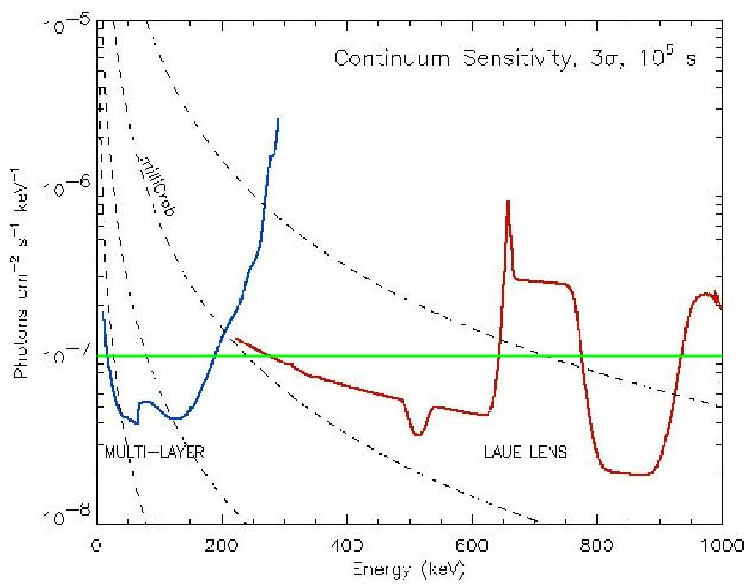}
\includegraphics*[width=8cm]{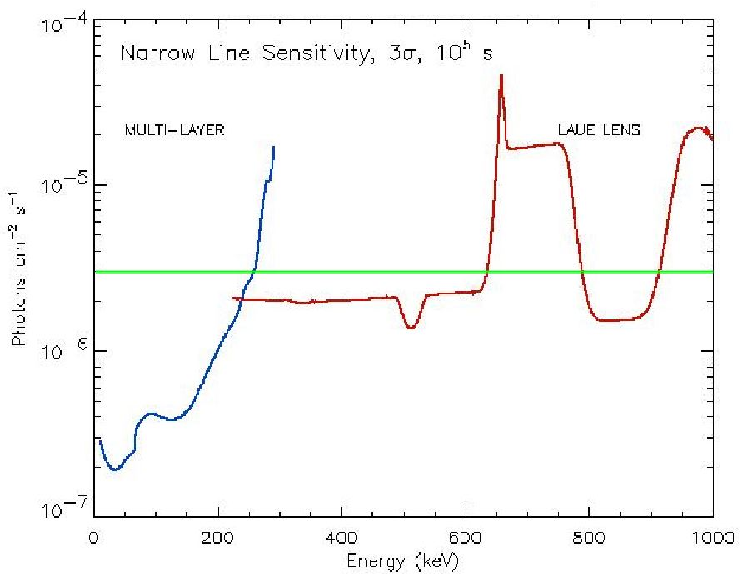}
\end{center}
\caption{\label{fig:performances}
Expected GRI performances.
Blue lines indicate the performances of the multilayer coated mirror, 
red lines give the performances of the Laue lens, and green lines 
indicate the mission requirements.
{\it Top-left:} Effective area as function of energy.
{\it Top-right:} Assumed instrumental background rate as function of 
energy.
{\it Bottom-left:} $3\sigma$ continuum detection sensitivity for an 
exposure time of 100 ks and $\Delta E/E=1/2$.
{\it Bottom-right:} $3\sigma$ narrow line detection sensitivity for an 
exposure time of 100 ks and $\Delta E/E=3\%$.
}
\end{figure}

The lens and mirror payload performances have been evaluated by dedicated 
simulation software based on ray tracing techniques. 
The critical performance parameters for these payloads are the expected 
effective area and focal spot characteristics as function of energy. 
For the detector payload, the most critical properties are the instrumental 
background and detection efficiency (which is limiting the sensitivity) and 
the spectral resolution (which is limiting the energy resolution). 
The instrumental background rates for GRI have been estimated using the 
measured background for the IBIS/ISGRI detector, reduced by a factor of 
4 as result of a better shielding and a lighter spacecraft mass (this 
shielding improvement is justified by the improvement achieved on 
INTEGRAL/SPI with the active BGO shield). 
The relevant background rates depend on the focal spot size on the detector. 
For a mirror HPD of 10 arcsec a focal spot size of $0.2$ cm$^2$ is expected, 
resulting in an active detector volume of $0.1$ cm$^3$. 
For a lens HPD of 30 arcsec a focal spot size of $3.2$ cm$^2$ is expected, 
resulting in an active detector volume of 21 cm$^3$.
The expected GRI performances are summarized in Fig.~\ref{fig:performances}.

\section{CONCLUSIONS}

The gamma-ray band presents a unique astronomical window that allows the 
study of the most energetic and most violent phenomena in our Universe.
With ESA's INTEGRAL observatory, an unprecedented global survey of 
the soft gamma-ray sky is currently performed, revealing hundreds
of sources of different kinds, new classes of objects, extraordinary views 
of antimatter annihilation in our Galaxy, and fingerprints of recent 
nucleosynthesis processes.
While INTEGRAL provides the longly awaited global overview over the soft 
gamma-ray sky, there is a growing need to perform deeper, more 
focused investigations of gamma-ray sources, comparable to the 
step that has been taken in X-rays by going from the EINSTEIN  
satellite to the more focused XMM-Newton observatory.

Technological advances in the past years in the domain of gamma-ray 
focusing using Laue diffraction techniques have paved the way towards 
a future gamma-ray mission, that will outreach past missions 
by large factors in sensitivity and angular resolution.
Complemented by a hard X-ray telescope based on a single-reflection 
multilayer coated concentrator, a broad-band energy coverage can be achieved 
that allows detailed studies of astrophysical sources at 
unprecedented sensitivity and angular resolution, from $\sim10$
keV up to at least 1 \MeV.
We have developed a mission concept, named the {\em Gamma-Ray Imager}, 
that combines both technologies in a promising new observatory.
Industry studies, conducted primarily by Thales Alenia Space and 
COSINE, have demonstrated that GRI is feasible and affordable within 
an ESA class M mission envelope.
GRI is composed of two spacecrafts that will be launched by a single Soyuz 
Fregate-2B launch from Kourou. 
The optics spacecraft (OSC) carries the multilayer coated single-reflection 
mirror and the Laue crystal lens that focus incoming gamma-rays onto a focal 
spot situated at a distance of 100~m. 
The detector spacecraft (DSC) carries the position sensitive detector, and 
will be placed in the focal spot to collect the concentrated gamma-rays. 
The estimated continuum sensitivity is better than $10^{-7}$ \feunit\
for a 100 ks exposure; the narrow line sensitivity will be better than 
$3 \times 10^{-6}$ \funit\ for the same integration time.
These sensitivities may be compared to those achieved today by 
INTEGRAL for the same exposure times:
$3 \times 10^{-6}$ \feunit\ at 100~keV (ISGRI) and
$1 \times 10^{-4}$ \funit\ at 847~keV (SPI).
With these unprecedented performances, a factor of more than 30 
better than existing mission, GRI will allow to study particle 
acceleration processes and explosion physics in unprecedented depth, 
providing essential clues on the intimate nature of the most violent 
and most energetic processes in the Universe.


\end{document}